\documentclass{article}
\usepackage{amssymb}
\newcommand{\bi}{\begin{itemize}}     
\newcommand{\ei}{\end{itemize}}
\newcommand{\bd}{\begin{description}} 
\newcommand{\ed}{\end{description}}
\newcommand{\bn}{\begin{enumerate}}   
\newcommand{\en}{\end{enumerate}}
\newcommand{\be}{\begin{equation}}
\newcommand{\ee}{\end{equation}}
\newcommand{\bc}{\begin{center}}
\newcommand{\ec}{\end{center}}
\newcommand{\ber}{\begin{eqnarray}}
\newcommand{\ear}{\end{eqnarray}}
\newcommand{\ba}{\begin{array}}
\newcommand{\ea}{\end{array}}

\newcommand{\al}{\alpha}
\newcommand{\bt}{\beta}
\newcommand{\bv}{\begin{verbatim}}
\newcommand{\bx}{\stackrel{2}{\nabla}}
\newcommand{\de}{\delta}

\newcommand{\ev}{\end{verbatim}}
\newcommand{\fr}{\frac}

\newcommand{\ga}{\gamma}

\newcommand{\la}{\lambda}

\newcommand{\Mu}{\check{\mu}}
\newcommand{\n}{\nonumber\\}
\newcommand{\na}{\nabla}

\newcommand{\ph}{\phi}
\newcommand{\rh}{\rho}

\newcommand{\si}{\sigma}
\newcommand{\ta}{\tau}

\newcommand{\te}{\theta}

\newcommand{\p}{\partial}

\begin{document}
\title{A Generalized Higgs Model.}
\author{Mark D. Roberts,\\
     {Department of Mathematics \& Statistics},
     {University of Surrey},  GU2 7XH,\\
            {mark.roberts@surrey.ac.uk},
     {http://www.maths.surrey.ac.uk/personal/st/Mark.Roberts}}
\maketitle
\bc Comment: Discussion of the principles involved added,  6 pages,
                      {\tt hep-th/9904080}
\ec
\begin{abstract}
The Higgs model is generalized so that in addition to the radial
Higgs field there are fields which correspond to the themasy and entropy.
The model is further generalized to include state and sign parameters.
A reduction to the standard Higgs model is given and how to break symmetry
using a non-zero VEV (vacuum expectation value) is shown.   A 'fluid rotation'
can be performed on the standard Higgs model to give a model dependent on
the entropy and themasy and with a constant mass.
\end{abstract}
{\small\tableofcontents}
\section{Introduction}
\subsection{The principles involved.}
One of the outstanding problems of particle physics is whether the Higgs'
field exists; these hypothetical fields are conjectured to make
non-Abelian gauge fields $A^\alpha_i$,  describe massive
particles as needed by the standard particle physics models.
There are \cite{bi:mdr89} alternatives to Higgs'
scalar fields which use fluids instead.
Apart from it being a good
idea to have alternative models, the reason for producing these
particular fluid alternatives can be addressed after outlining the
principles below:
\bd
\item[The principle that:  ''All fundamental fields are gauge fields''.]
The success of non-Abelian gauge
vector fields in describing fundamental interactions suggests the
principle that ``{\sc All fundamental field are gauge fields.}''
Taken at face value this implies that the Higgs
fields used in symmetry
breaking should also be a gauge field: as it is a scalar field at
first sight this seems impossible. A way around this is to have a
collection of scalar fields with the properties that:
{\it firstly} they encompass the standard
Higgs mechanism,  {\it secondly}
they form a gauge system. This can be realized by using the scalar
field description of a perfect fluid, and then ``charging'' it via
$\partial_a\rightarrow D_a=\partial_a+ieA_a$, as described below.
\item[The: "Extreme principle of equivalence."]
The extreme principle of equivalence is that "{\sc There is only
one concept of mass in physics.}" \cite{bi:mdr89},
and this extended principle requires the
equivalence of these masses to all other masses, including those
generated by symmetry breaking mechanisms.
\item[The: "Redundancy of {\it ad hoc} procedures."]
This
principle is that "{\sc Any principled explanation (perhaps not
only scientific) of anything is preferable to an {\it ad hoc}
calculational procedure which produces an equivalent or inferior
result.}"  It leaves open what to choose if the {\it ad hoc}
calculational procedure produces a superior result, or how to
address the quality of the result. In the present case
Higgs' scalar fields are {\it ad hoc}
whereas fluids might arise from the statistical properties of
the non-Abelian gauge fields.
\ed
\subsection{Application of these principle to symmetry breaking.}
\paragraph{Re-writing procedure.}
There are well known techniques by which stresses involving scalar
fields can be rewritten as fluids, so that it is possible to
re-write the standard model with fluids instead of
Higgs scalars.   My {\em first}
attempt at using fluids for symmetry breaking
\cite{bi:mdr89} was essentially to
deploy these rewriting procedures to convert the scalar fields to
fluids;  the drawback of this approach is the fluids that result
are somewhat unphysical,  however an advantage is that symmetry
breaking occurs with a change of state of the fluid.
\paragraph{Velocity potential method.}
My {\em second} attempt
\cite{bi:mdr97} used the decomposition of a perfect fluid vector into
several scalar parts called vector potentials and identifying one
of these with the radial Higgs scalar; thus the theory is a generalization of the standard
theory, so that any experimental verification of the standard
theory does not negate it. The gauge fields are then introduced by
using the usual covariant substitutions for the partial
derivatives of the scalars,  for example
$\phi_a=\partial_a\phi\rightarrow\nabla_a\phi=\partial_a\phi+ie\phi A_a$,
and calling the resulting fluid the
{\sc covariantly interacting fluid}.
This results in an elegant extension of
standard Higgs symmetry
breaking,  but with some additional parameters present. Previous
work on the vector potentials shows that some of these have a
thermodynamic interpretation, it is hoped that this is inherited
in the fluid symmetry breaking models.   The additional parameters
can be partially studied with the help of an explicit
Lagrangian,  see below.
Both of my approaches have been so far restricted to abelian gauge fields.

\subsection{The lagrangian formulation of fluids.}
A perfect fluid has a Lagrangian formulation in which the Lagrangian
is the pressure $p$.    Variation is achieved by using the first law of
thermodynamics
\be
dp=n~dh-nT~ds,
\label{eq:1.1}
\ee
where $n$ is the particle number,  $T$ is the temperature,  $s$ is the entropy,
and $h$ the enthalpy.   In four dimensions a vector can be decomposed into four
scalars,  however the five scalar decomposition
\be
hV_\mu=W_\mu=\si_\mu+\sum_{i}\te^{(i)}s_{(i)\mu},~~~~~~~~V_\rh V^\rh=-1,
\label{eq:1.2}
\ee
$(i)=1,2$  is often used,  because for $i=1$,  $s$ and $\te=\int T~d\ta$
have interpretation, Roberts (1997) \cite{bi:mdr97}, as the entropy and the
thermasy respectively.   From now on the index $(i)$ is suppressed as it is
straightforward to reinstate.   Replacing the first law with
$dp=-V_\rh dW^\rh-nT~ds$,  variation gives the familiar
stress and scalar evolution equations.
Previously,  Roberts (1997) \cite{bi:mdr97},   all
partial derivatives in the above have been replaced with vector covariant derivatives
\be
D_\mu=\p_\mu+ie~A_\mu,
\label{eq:1.3}
\ee
to obtain a generalization of scalar electrodynamics called fluid
electrodynamics.   For many interacting fluids the interaction
terms can be disregarded,  Anile (1990) \cite{bi:anile},
however here all interaction terms are kept.   Quantization of these
models is not looked at here although fluids can be quantized,
Roberts (1999) \cite{bi:mdr99}.
Here instead of using the first law for variation a specific
explicit Lagrangian is assumed.   This fixes the equation of state.
Since previous work three things are approached differently.
The {\it first} is the best
conventions for the scalar decomposition are \ref{eq:1.2}.
This is because the spacetime index is put on $s$ as this will allow easier
generalization to second order non-equilibrium thermodynamics,
Israel and Stewart (1979) \cite{bi:IS}.  The $+$ convention is used
for $\te$ and $s$ in \ref{eq:1.2} (rather than $-\te_\mu s$).
The {\it second} is that the fluid remains isentropic after charging.
The {\it third} is that two vector normalization conditions are required
after charging.   Equivalences between fluids and scalar fields were first
studies by Tabensky and Taub (1973) \cite{bi:TT}.
Thermodynamical quantities can be introduced into the standard Higg's model via
the partition function Kapusta (1989) \cite{bi:kapusta}.
Clearly the entropy,  temperature and so on cannot occur from both
the vector \ref{eq:1.2} and the partition function,  perhaps
a suitable partition function for the present model might allow both to be
identified.
An approach which dispenses with Higgs fields is that of
Nicholson and Kennedy (2000) \cite{NK},
another approach which involves gravity
is that of Kakushadze and Langfelder (2000) \cite{KL}.
\section{Explicit Lagrangians}
Consider the Lagrangian
\be
L(q,q_\rh)=\bt(-W_\rh W^\rh)^{r}-Q(q),
\label{eq:2.1}
\ee
where $r$ is called the equation of state parameter
(see equation \ref{eq:2.6}),
$\bt$ is called the sign parameter and $Q$ is the potential.
Varying with respect to the metric and then assuming the stress is of the
form of a perfect fluid implies
\be
-2(-1)^{r}\bt rW_\rh^{2r-2}W_\mu W_\nu=(p+\mu)V_\mu V_\nu.
\label{eq:2.2}
\ee
Normalization $V_\rh^{2}=-1$ and $hV_\mu=W_\mu$ implies
\be
W_\rh^{2r}=(-1)^{r}h^{2r}.
\label{eq:2.3}
\ee
Recall that $L=p$,  together with the above three equations this gives
\be
p=\bt h-Q,~~~
\mu=(2r-1)\bt h^{2r}+Q,~~~
n=2\bt rh^{2r-1}.
\label{eq:2.4}
\ee
Varying $L$ with respect to $\si$, $\te$ and $s$ respectively
\be
(nV^\rh)_\rh-Q_{\si}= 0,~~~
-n\dot{s}-Q_{\te}=0,~~~
(n\te V^\rh)_\rh-Q_{\si}=0,
\label{eq:2.5}
\ee
This is a perfect fluid;  when $Q=0$ it has the  $\ga$-equation of state
\be
1<\ga=\fr{2r}{2r-1}<2,~~~
\fr{\ga}{2\ga-2}=r<\fr{1}{2}.
\label{eq:2.6}
\ee
Particular cases are $r=1$ which implies $\ga=2$ which is the equation of
state for coherent radiation;  and $r=2$ which implies $\ga=\fr{4}{3}$
which is the equation of state for incoherent radiation.
The canonical momenta are
\be
\Pi^{\si}=-n,~~~
\Pi^{\te}=0,~~~
\Pi^{s}=n\te.
\label{eq:2.7}
\ee
The constrained Hamiltonian is
\be
H_{\la}=\Pi^{\si}(\dot{\si}+\te\dot{s})
           +\la^{1}(\Pi^{s}-\te\Pi^{\si})
           +\la^{2}\Pi^{\te}-L,
\label{eq:2.8}
\ee
where $\la^{1}$ and $\la^{2}$ are the Lagrange multipliers.
The momenta and Hamiltonian are the same as in the general case.
In the general case the Euler equations and the canonical stress vanish
identically;  however for explicit Lagrangians the Euler equations are the
same as \ref{eq:2.5} and the canonical stress is the same as the stress.
\section{Charged Explicit Lagrangian.}
All partial derivatives are replaced by vector covariant derivative
\ref{eq:1.1}.   Capital letters are used for the new quantities and
small letters for the uncharged quantities.   The Lagrangian becomes
\be
L\rightarrow L=\bt(-W_\rh W^{*\rh})^{r}-Q(q)-\fr{1}{4}F^{2},
\label{eq:3.1}
\ee
where "*" denotes complex conjugate.   The vector field becomes
\be
W_\mu=\si_\mu+\te s_\mu+i(\si+\te s)eA_\mu=w_\mu+i\bar{e} A_\mu,
\label{eq:3.2}
\ee
where $\bar{e}\equiv(\si+\te s)e$.   Under the global transformations
\be
\si\rightarrow exp(-ie\la)\si,~~~
s\rightarrow exp(-ie\la)s,~~~
\te\rightarrow\te,~~~
A_\mu\rightarrow A_\mu+\la_\mu,
\label{eq:3.3}
\ee
the vector field changes to $W_\mu\rightarrow exp(-ie\la)W_\mu$
so that $W_\rh W^{*\rh}$ is invariant,
implying that the Lagrangian is invariant.
One can choose that $A_\mu=0$ and then these transformations are local.
The N\"other current is
\be
J^\mu=\bt r(-W_\rh W^{*\rh})^{r-1}\left\{
-i(\bar{e}^*w^\mu-\bar{e}w^{*\mu})+2\bar{e}\bar{e}^*A^\mu\right\}.
\label{3.3b}
\ee
One can introduce two normalization conditions
\be
V_\mu V^{*\mu}=-1,~~~
v_\mu v^\mu=-1,
\label{eq:3.5}
\ee
which imply
\be
H^{2}=h^{2}-e^{2}A_\rh^{2},
\label{eq:3.6}
\ee
Proceeding as before
\be
P=\bt H^{2r}-Q-\fr{1}{4}F^{2},~~~
\Mu=\bt(2r-1)H^{2r}+Q+\fr{1}{4}F^{2},~~~
N=2\bt rH^{2r-1},
\label{eq:3.7}
\ee
and the canonical momenta are
\be
\Pi^{\si}=-N,~~~
\Pi^{\te}=0,~~~
\Pi^{s}=-N\te,~~~
\Pi^{A_\mu}=0,
\label{eq:3.8}
\ee
the last of which is surprising;  because not all the components of
$\Pi^{A_\mu}$ vanish in usual gauged scalar electrodynamics.   The stress is
\ber
T_{\mu\nu}&=&NH(V_\mu V^*_\nu+V^*_\mu V_\nu)+Pg_{\mu\nu},\n
      &=& \fr{N}{H}(w_\mu w_\nu+e^2A_\mu A_\nu)+Pg_{\mu\nu},\n
     T&=&3P-\Mu,\n
v^\al v^\bt T_{\al\bt}&=&\mu+e\fr{N}{H}\left(A_\al^{2}+(v^\al A_\al)^2\right).
\label{eq:3.9}
\ear
The conservation law is
\be
T_{.\mu.;\bt}^\bt=w_\mu Q_{\si}
            +N\fr{h}{H}\dot{w}_\mu
            +e^{2}(\fr{N}{H}A_\mu A^\bt)_\bt
            +P_\mu.
\label{eq:3.10}
\ee
The constrained Hamiltonian is
\be
H_{\la}=\Pi^{\si}(\dot{\si}+\te\dot{s})+\la^{1}(\Pi^{s}-\te\Pi^{\si})
                                       +\la^{2}\Pi^{\te}
                                       +\la^{3}\Pi^{A_{a}}-L,
\label{eq:3.11}
\ee
Variation with respect to the scalar and vector fields gives
\ber
\de\si&:&Q_{\si}=(N\fr{h}{H}v^\al)_\al,\n
\de\te&:&-Q_{\te}=N\fr{h}{H}\dot{s},\n
\de s&:&Q_{s}=(N\fr{h}{H}v^\al)_\al,\n
\de A_\mu&:&F_{\mu;\bt}^\bt=e^2\fr{N}{H}A_\mu.
\label{eq:3.12}
\ear
Note that if one starts with the electrodynamical Lagrangian ${\cal{L}}=-F^2/4$
and tries to implement a substitution scheme from this via
$A_\mu\rightarrow A_\mu+kW_\mu$ one cannot recover the above type of charged
fluids because $F^2\rightarrow F^2+$ second derivatives in the scalar fields,
which do not occur in the above.   Thus there is no mirror mechanism.
\section{Comparison with the Higgs' Model}
To recover scalar electrodynamics first note that
\be
W_\al W^{*\al}=w_\al^2+\bar{e}^2A_\al^2.
\label{eq:4.1}
\ee
Now set the state parameter $r=1$,  the sign parameter $\bt=-1$,
and the potential $Q(q)=V(\rh^{2})$ so that
\be
L=w_\al^2+\bar{e}^2A_\al^2-V(\rh^2)-\fr{1}{4}F^{2},
\label{eq:4.2}
\ee
which has no cross term $w_\al A^\al$.
Now set $\si=\rh$,  $\te=s= 0$.  Changing the gauge $A'_\mu=\nu_\mu+A_\mu$ and
dropping the prime gives {\it scalar electrodynamics} in radial form.
This has lagrangian,  stress and equations of motion
\ber
L&=&\rh_\al^2+(\hat{\na}_\al\nu)^{2}-V(\rh^{2})-\fr{1}{4}F^{2},\n
\hat{\na}_\mu\nu&=&\rh(\nu_\mu+eA_\mu),\n
T_{\mu\nu}&=&2\ph_\mu\ph_\nu+2\hat{\na}_\mu\nu\hat{\na}_\nu\nu
                        +F_{\mu\ga}F_{\nu.}^{.\ga}+g_{\mu\nu}L,\n
F^{\mu\bt}_{..\bt}&+&2e\rh\hat{\na}^\mu\nu=0,\n
2[\bx&+&(\nu_\bt+eA_\bt)^{2}+V']\rh=0,\n
2[\bx\nu&+&eA^\bt_\bt]=0,
\label{eq:4.3}
\ear
respectively.   Giving the Higgs field a non-zero expectation value
$<0|\rh|0>= a$.   $\rh$ transfers to
\be
\rh \rightarrow \rh ' = \rh +a,
\label{eq:4.4}
\ee
and the lagrangian becomes
\be
L=\rh_\bt^{2}+(\rh+a)^2e^2A_\bt^2-V\left((\rh+a)^{2}\right)
             -\fr{1}{4}F^{2},
\label{eq:4.5}
\ee
the term $a^2e^2A_\bt^2$ is a constant mass term.

For a fluid one can perform a 'scalar fluid rotation'
\be
\si \rightarrow \si '=\si+\te s.
\label{eq:4.6}
\ee
Starting with scalar electrodynamics and identifying $\rh$ with $\fr{\si}{e}$
does not give one of the above fluids as gradients in $\te$ appear.
Alternatively there is the 'vector fluid rotation'
\be
\si_\mu=\rh_\mu\rightarrow w_\mu.
\label{eq:4.7}
\ee
Starting with scalar electrodynamics and performing this rotation and replacing $\rh$
with $\fr{\si}{e}$ gives a fluid of the above type.
One can then assume a VEV $<0|\fr{\si}{e}|0>$  to
generate a constant mass term as above,   the difference being that some
account of thermodynamics has been achieved.
\section{Acknowledgements}
I would like to thank Prof.T.W.B.Kibble
for his interest in this work.


\begin{thebibliography}{99}
\bibitem{bi:vandantzig}
van Dantzig,D.(1939)\\
On the Phenomenological Thermodynamics of Moving Matter.\\
{\it Physica},{\bf 6}(1939)673-704.

\bibitem{bi:mdr89}
Roberts,Mark D.\\
Symmetry Breaking using Fluids and the Extreme Equivalence Principle.\\
{\it Hadronic J.}{\bf 12}(1989)93-99.

\bibitem{bi:mdr97}
Roberts,Mark D.\\
Fluid Symmetry Breaking II:  Velocity Potential Method.\\
{\tt hep-th/9904079}
{\it Hadronic J.}{\bf 20}(1997)73-84.

\bibitem{bi:anile}
Anile,A.M.(1990)\\
Relativistic Fluids and Magneto-Fluids:\\
with Applications in Astrophysics and Plasma Physics.\\
Cambridge University Press Monographs in Mathematical Physics.

\bibitem{bi:mdr99}
Roberts,Mark D.\\
The Quantum Commutator of a Perfect Fluid.\\
{\tt gr-qc/9810089}
{\it Mathematical Physics,  Geometry and Analysis.}{\bf 1}(1999)367-373.

\bibitem{bi:IS}
Israel,W. \& Stewart,J.M.\\
Transient Relativistic Thermodynamics and Kinetic Theory.\\
{\it Ann.Phys.(N.Y.)}{\bf 118}(1979)341.

\bibitem{bi:TT}
Tabensky,R.  \&  Taub,A.H.\\
Plane symmetric self-gravitating fluids\\
with pressure equal to energy density.\\
{\sl Math.Rev.{\bf 47}\#10022}
{\it Commun.Math.Phys.}{\bf 29}(1973)61.

\bibitem{bi:kapusta}
Kapusta,J.(1989)\\
Finite-Temperature Field Theory.\\
Cambridge University Press,  Cambridge.\\
{\sl Math.Rev.92e:81003}

\bibitem{NK}
Nicholson,  Angus F. \& Kennedy,  Dallas C.\\
Electroweak Theory Without Higgs Bosons.\\
{\tt hep-ph/9706471}
{\it Int.J.Mod.Phys.}{\bf A15}(2000)1497.

\bibitem{KL}
Kakushadze, Zurab \& Langfelder, Peter\\
Gravitational Higgs Mechanism.\\
{\tt hep-th/0011245}
{\it Mod.Phys.Lett.}{\bf A15}(2000)2265-2280.

\end{thebibliography}
\end{document}